\documentclass{article}
\usepackage{spconf,amsmath,graphicx}
\usepackage{multirow}
\usepackage{booktabs}
\usepackage{amssymb}
\usepackage{bm}
\usepackage{float}
\usepackage[export]{adjustbox}

\usepackage{enumitem}
\usepackage{multirow}
\usepackage{makecell}
\usepackage{tabularx}
\usepackage{booktabs}
\usepackage{xcolor}

\setlist{nosep, leftmargin=14pt}
\usepackage[left=1.62cm,right=1.62cm,top=1.9cm]{geometry}

\title{BENCHMARKING ULTRASOUND FOUNDATION MODELS\\
 FOR FETAL PLANE CLASSIFICATION\\ }

\name{
Leya~Barrientos$^{1}$,  
Yuexi Du$^{2}$, 
Nicha~C.~Dvornek$^{1,2}$
}

\address{
$^{1}$ Radiology \& Biomedical Imaging, Yale School of Medicine, USA\\ 
$^{2}$ Department of Biomedical Engineering, Yale University, USA
}
\begin{document}
%\ninept
%
\maketitle

\begin{abstract}
Ultrasound is widely used in obstetric care due to its safety, accessibility, and real-time imaging. However, interpretation remains operator-dependent and susceptible to noise and artifacts. Deep learning models have shown strong performance to solve these problem, but they typically require large annotated datasets that are difficult to obtain in clinical ultrasound. Foundation models (FMs) offer an alternative, using a large number of ultrasound images to learn transferable representations that can generalize with limited labeled data. This work presents a comprehensive benchmark of ultrasound-specific FMs for fetal plane classification. We evaluated four ultrasound FMs (USFM, MOFO, UltraSAM, FetalCLIP) against two CNN baselines (ResNet50, EfficientNet-V2) and a ViT (DINOv3) pretrained on natural images. We trained all models under two complementary settings: full fine-tuning and linear probing with a frozen encoder. All models were trained using 5-fold patient-level cross-validation on a Spanish fetal ultrasound dataset and tested on both in-domain data and an external African cohort to assess cross-population generalization. We found that FetalCLIP achieved the best results in the linear probing setting (F1 = 0.9261 for in-domain, F1 = 0.9731 for out-of-domain), while USFM performed best in the full fine-tuning setting (F1 = 0.9476 for in-domain, F1 = 0.9515 for out-of-domain). MOFO and UltraSAM degraded most in both settings, underperforming natural image pretrained models in some cases. These findings highlight how the choice of pretrained model strongly affects fetal plane classification performance, since different pretraining objectives lead to different levels of transferability.
%These findings highlight the promise of ultrasound-specific self-supervised pretraining, particularly masked image modeling, for robust fetal plane classification.
%FM models are the best, it's not only the size. MIM and CLIP.

\end{abstract}
% %
\begin{keywords}
 Ultrasound, Fetal plane classification, Foundation models, Domain generalization
\end{keywords}
% %

\section{Introduction}
\label{sec:intro}

Ultrasound is one of the most accessible, affordable, and safe imaging modalities in clinical practice, playing a critical role in routine screening, interventional procedures, and especially in fetal care \cite{wells2006ultrasound}. However, ultrasound images inherently suffer from a low signal-to-noise ratio, while also being highly operator-dependent.

Recent advances in artificial intelligence \cite{wang2019artificial}  have shown promise for improving ultrasound interpretation and reducing operator dependence. However, traditional deep learning (DL) models require large, carefully annotated datasets, which are difficult to obtain in ultrasound due to expert labeling demands and variability in acquisition. Therefore, many existing DL models struggle to generalize across different clinical settings.

%It can range from natural images, US images, target-specific images (fetal CLIP) 

Foundation models (FMs) \cite{schneider2024foundation} have emerged as a promising solution to the limitations of conventional deep learning in ultrasound imaging. These models are large, pre-trained on broad and diverse ultrasound datasets, and can be efficiently adapted for downstream applications by fine-tuning them with relatively little labeled data.

These FMs are especially valuable for fetal plane classification, a task that ensures accurate biometric and anatomical assessment in obstetric ultrasound \cite{salomon2011practice}. Because each ultrasound examination contains many images and only a subset corresponds to standardized diagnostic planes, manual selection is slow and prone to error. An automatic classifier can efficiently identify the fetal planes, improve consistency, and reduce the workload for sonographers and fetal medicine specialists.

In this paper, we present a comprehensive benchmarking study that evaluates multiple state-of-the-art ultrasound foundation models and conventional architectures pretrained on natural images on fetal plane classification. By assessing their performance and generalizability across diverse datasets, our work aims to provide insights into the strengths, limitations, and practical readiness of recent ultrasound foundation models for real-world clinical deployment.

\section{Methods}

\begin{figure*}[!t]
    \centering
    \includegraphics[width=0.88\textwidth]{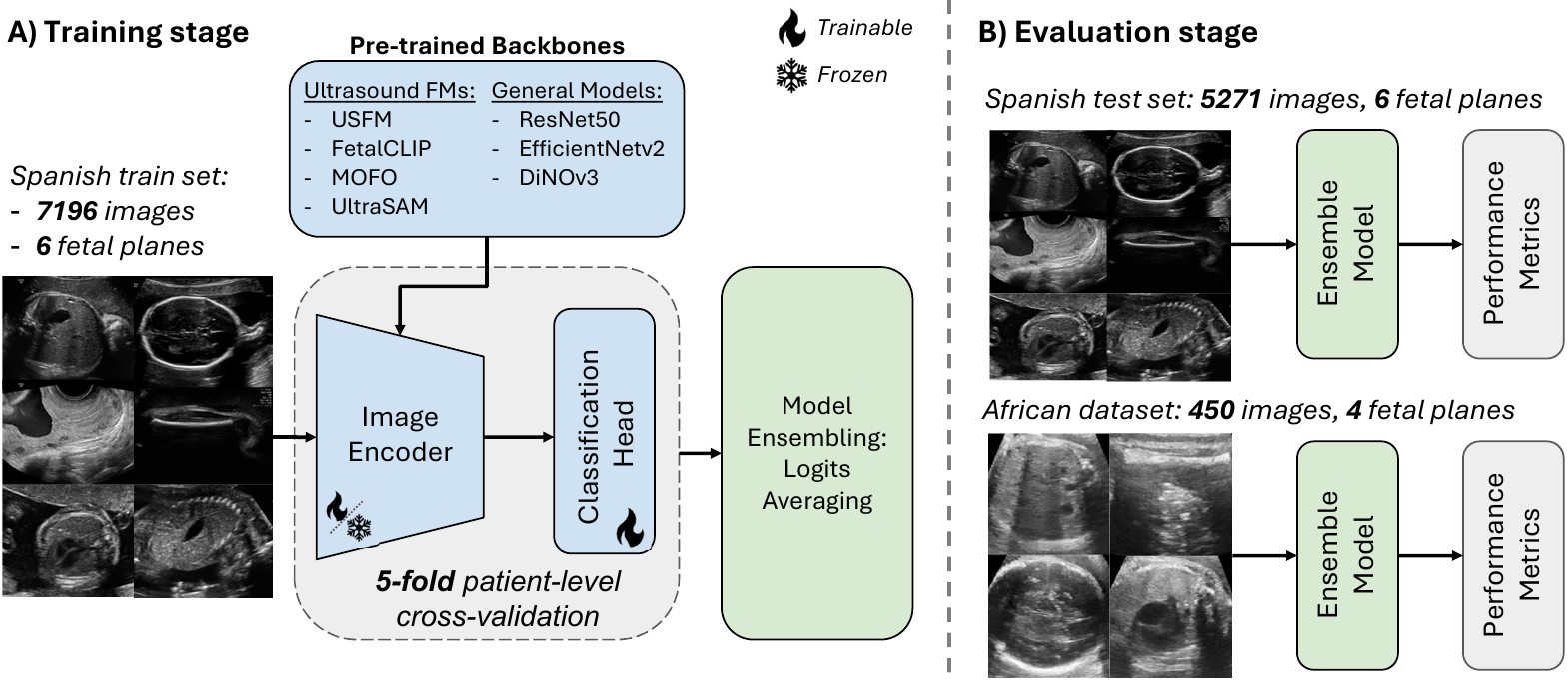}
    \vspace{-2mm}
    \caption{Workflow of the proposed (A) training and (B) evaluation pipeline. Multiple backbones are trained with 5-fold patient-level cross-validation, and ensemble inference is performed using logit averaging.}
    \vspace{-4mm}
    \label{methodology}
\end{figure*}

We benchmark four ultrasound foundation models against natural image models to assess the impact of domain-specific pretraining in fetal plane classification. We divided the experiment into two stages: training and evaluation, as described in Fig.~\ref{methodology}.

\subsection{Datasets}

%We employed two publicly available ultrasound datasets of standard fetal planes. 

\noindent\textbf{Spanish Dataset:}~  A public dataset collected from a Spanish population comprises 12,400 ultrasound images categorized into six fetal plane classes: abdomen, brain, femur, thorax, maternal cervix, and others \cite{burgos2020evaluation}. The class distribution is significantly imbalanced: fetal abdomen (5.7\%), femur (8.4\%), thorax (13.9\%), cervix (13.1\%), brain (24.9\%), and other (33.9\%). We used the official data split \cite{burgos2020evaluation}, with 7,129 images in the training set and 5,271 images in the test set, each containing data from 896 patients. 

\noindent\textbf{African Dataset:}~ A second public dataset contains 450 images from 100 patients acquired from an African population \cite{sendra2023generalisability}. It includes four fetal standard planes: abdomen, brain, femur, and thorax. The class distribution is relatively balanced, consisting of 125 abdomen (27.8\%), 125 brain (27.8\%), 125 femur (27.8\%), and 75 thorax (16.7\%) images. 

\subsection{Model architectures} 
%\todo{Subsections for US foundation model, natural image foundation model, classification head}
\hspace*{1em}
\noindent\textbf{Ultrasound Foundation Models:}~ We evaluated four state-of-the-art ultrasound foundation models, selected based on their public availability. The first model is \textbf{USFM} (UltraSound Foundation Model)~\cite{jiao2024usfm}, a self-supervised approach employing a Vision Transformer (ViT) encoder trained with Masked Image Modeling (MIM) on 2,187,915 ultrasound images. The second model is \textbf{MOFO} (Multi-Organ FOundation)~\cite{chen2024multi}, which combines a CSWin-ViT encoder with a CNN decoder and was trained on 7,039 ultrasound images. The third model is \textbf{UltraSAM}~\cite{meyer2025ultrasam}, an ultrasound-adapted variant of the Segment Anything Model (SAM \cite{kirillov2023segment}), which also uses a ViT encoder and was trained on 280,000 ultrasound images \cite{meyer2025ultrasam}. Lastly, we evaluate \textbf{FetalCLIP}~\cite{maani2025fetalclip}, an ultrasound ViT model pretrained with contrastive language-image learning on 207,943 paired fetal ultrasound images and captions. For the segmentation-oriented FMs, we only use their image encoder as the backbone for downstream fine-tuning.

\noindent\textbf{Natural Image Models:}~ To assess the benefit of domain-specific pretraining, we also evaluated models pretrained on natural images. For CNN-based architectures, we selected \textbf{ResNet50} \cite{he2016deep} and \textbf{EfficientNet-V2} \cite{tan2021efficientnetv2} pre-trained on ImageNet-1k~\cite{deng2009imagenet}. For Transformer-based architectures, we used \textbf{DINOv3}, a self-supervised ViT model~\cite{simeoni2025dinov3}. Since all four ultrasound foundation models are Transformer-based, this comparison allows us to examine whether a Transformer model pretrained on billions of natural images can match or surpass the domain-specific model trained on a much smaller scale.

This selection strategy enabled fair and comprehensive benchmarking across different architectures while maintaining practical relevance for potential clinical deployment.

\noindent\textbf{Fetal Plane Classification Model:} A linear classification head was appended to the pretrained encoder of each backbone model to predict the fetal plane classification. 

\begin{table*}[htbp]
\centering
\caption{Full fine-tuning fetal plane classification results across datasets and model types.}
\label{tab:benchmark_results}
\resizebox{0.9\textwidth}{!}
{
\begin{tabularx}{\textwidth}{p{3cm} p{2cm} p{2.6cm} *{6}{>{\centering\arraybackslash}X}}
\toprule
\textbf{Dataset} & \textbf{Type} & \textbf{Model} & \textbf{ACC} & \textbf{AUC} & \textbf{PREC} & \textbf{RECALL} & \textbf{F1} & \textbf{SPEC} \\
\midrule

% ---------------- SPANISH -----------------
\multirow{7}{*}{\makecell{\textbf{Spanish Population}\\(In-domain)}}
 & \multirow{2}{*}{CNN-based}
 & ResNet50        & 0.9463 & 0.9954 & 0.9323 & 0.9495 & 0.9404 & 0.9889 \\
 &                 & EfficientNetV2 & 0.9421 & 0.9957 & 0.9223 & 0.9562 & 0.9371 & 0.9886 \\
\cmidrule(lr){2-9}
 & ViT-based
 & DINOv3          & 0.7366 & 0.9638 & 0.8056 & 0.5977 & 0.6058 & 0.9374 \\
\cmidrule(lr){2-9}
 & \multirow{4}{*}{FM-based}
 & UltraSAM        & 0.7417 & 0.9281 & 0.7198 & 0.6234 & 0.6247 & 0.9429 \\
 &                 & MOFO            & 0.9185 & 0.9907 & 0.9051 & 0.9053 & 0.9049 & 0.9827 \\
 &                 & USFM            & \textbf{0.9521} & \textbf{0.9962} & \textbf{0.9395} & \textbf{0.9567} & \textbf{0.9476} & \textbf{0.9901} \\
 &                 & FetalCLIP       & 0.8446 & 0.9080 & 0.7037 & 0.7471 & 0.7270 & 0.9710 \\
\midrule

% ---------------- AFRICAN -----------------
\multirow{7}{*}{\makecell{\textbf{African Population}\\(Out-of-domain)}}
 & \multirow{2}{*}{CNN-based}
 & ResNet50        & 0.9156 & 0.9811 & 0.9173 & 0.8947 & 0.9015 & 0.9713 \\
 &                 & EfficientNetV2 & 0.9289 & 0.9854 & 0.9240 & 0.9187 & 0.9199 & 0.9763 \\
\cmidrule(lr){2-9}
 & ViT-based
 & DINOv3          & 0.6311 & 0.8831 & 0.7575 & 0.5760 & 0.5297 & 0.8723 \\
\cmidrule(lr){2-9}
 & \multirow{4}{*}{FM-based}
 & UltraSAM        & 0.6156 & 0.8277 & 0.6388 & 0.6313 & 0.6022 & 0.8775 \\
 &                 & MOFO            & 0.8644 & 0.9481 & 0.8733 & 0.8393 & 0.8472 & 0.9537 \\
 &                 & USFM            & \textbf{0.9556} & \textbf{0.9938} & \textbf{0.9564} & \textbf{0.9480} & \textbf{0.9515} & \textbf{0.9849} \\
 &                 & FetalCLIP       & 0.6067 & 0.8612 & 0.5611 & 0.6460 & 0.5501 & 0.8814 \\
\bottomrule

\end{tabularx}
}
\vspace{-2mm}
\end{table*}

\subsection{Training Stage} The training framework is illustrated in Fig.~\ref{methodology}A. The Spanish training set was used to train all classification models to predict all six fetal anatomical planes. We used a 5-fold patient-level cross-validation strategy, where in each iteration, four folds were used for training and the remaining fold for validation. The model checkpoint with the highest validation accuracy from each fold was saved. A final model per backbone was generated using logit averaging. Specifically, the raw logits from the five cross-validation models were averaged to form an ensemble predictor.

We trained all models under two complementary settings: full fine-tuning and linear probing. In the full fine-tuning setup, all parameters of the backbone and classifier head were updated. In contrast, the linear probing setup froze the entire pretrained backbone and trained only the lightweight classification head.

\subsection{Training Details}
All models were implemented in the PyTorch framework. Ultrasound images were resized to 224×224 pixels. For models trained on grayscale inputs, images were first converted to a single-channel grayscale representation. For architectures requiring three-channel input, the grayscale image was replicated across channels. CNN-based models, USFM, and MOFO were normalized using a mean of 0.5 and standard deviation of 0.5. UltraSAM followed Meta’s default SAM normalization. For DINOv3, preprocessing (including  RGB conversion and normalization) was handled using the Hugging Face AutoImageProcessor to ensure compatibility with the pretrained backbone.

For all models except DINOv3, data augmentation was applied using a set of standard image transformations, including random horizontal flips, random rotations up to ±15°, and random translations of up to 10\% in both directions, each executed with a probability of 0.5. For DINOv3, we relied on the augmentation strategy from the original pretraining.

Each model was trained using cross-entropy loss for 20 epochs with a batch size of 32. Following optimizer recommendations for each architecture, CNN-based models, MOFO, and USFM were optimized using Adam, while UltraSAM and DINOv3 were trained with AdamW. All models used an initial learning rate of $1\times10^{-4}$, reduced by a factor of 0.1 every 7 epochs, and a weight decay of $1\times10^{-4}$ for regularization. 

\subsection{Evaluation Stage}

To assess model performance, we conducted two evaluation experiments, as shown in Fig.~\ref{methodology}B. First, we performed in-domain evaluation, where each model was tested on the held-out test set of the Spanish dataset. Then, to evaluate out-of-domain generalizability, we performed external testing on the African dataset, which includes only four of the six fetal plane categories present during training. To preserve a consistent six-dimensional probability output, the logits corresponding to the two absent classes were masked and set to zero prior to softmax normalization. 

\begin{table*}[htbp]
\centering
\caption{Linear probing fetal plane classification results across datasets and model types.}
\label{tab:benchmark_results_frozen}
\resizebox{0.9\textwidth}{!}
{
\begin{tabularx}{\textwidth}{p{3cm} p{2cm} p{2.6cm} *{6}{>{\centering\arraybackslash}X}}
\toprule
\textbf{Dataset} & \textbf{Type} & \textbf{Model} & \textbf{ACC} & \textbf{AUC} & \textbf{PREC} & \textbf{RECALL} & \textbf{F1} & \textbf{SPEC} \\
\midrule

% ---------------- SPANISH (FROZEN) -----------------
\multirow{7}{*}{\makecell{\textbf{Spanish Population}\\(In-domain)}}
 & \multirow{2}{*}{CNN-based}
 & ResNet50        & 0.8210 & 0.9700 & 0.8367 & 0.7146 & 0.7320 & 0.9595 \\
 &                 & EfficientNetV2 & 0.8378 & 0.9738 & 0.8255 & 0.7819 & 0.7946 & 0.9648 \\
\cmidrule(lr){2-9}
 & ViT-based
 & DINOv3          & 0.8366 & 0.9780 & 0.8571 & 0.7385 & 0.7606 & 0.9633 \\
\cmidrule(lr){2-9}
 & \multirow{4}{*}{FM-based}
 & UltraSAM        & 0.2824 & 0.5977 & 0.0868 & 0.1760 & 0.1108 & 0.8332 \\
 &                 & MOFO            & 0.6219 & 0.8497 & 0.4625 & 0.4637 & 0.4489 & 0.9156 \\
 &                 & USFM            & 0.9117 & 0.9883 & 0.8981 & 0.8925 & 0.8989 & 0.9804 \\
 &                 & FetalCLIP       & \textbf{0.9321} & \textbf{0.9938} & \textbf{0.9223} & \textbf{0.9305} & \textbf{0.9261} & \textbf{0.9857} \\
\midrule

% ---------------- AFRICAN (FROZEN) -----------------
\multirow{7}{*}{\makecell{\textbf{African Population}\\(Out-of-domain)}}
 & \multirow{2}{*}{CNN-based}
 & ResNet50        & 0.5533 & 0.8583 & 0.6461 & 0.5033 & 0.4747 & 0.8457 \\
 &                 & EfficientNetV2 & 0.6311 & 0.8802 & 0.6106 & 0.5813 & 0.5594 & 0.8745 \\
\cmidrule(lr){2-9}
 & ViT-based
 & DINOv3          & 0.5511 & 0.9132 & 0.7252 & 0.5010 & 0.4496 & 0.8448 \\
\cmidrule(lr){2-9}
 & \multirow{4}{*}{FM-based}
 & UltraSAM        & 0.2822 & 0.5238 & 0.1769 & 0.2566 & 0.1299 & 0.7522 \\
 &                 & MOFO            & 0.5333 & 0.7704 & 0.5928 & 0.5333 & 0.5290 & 0.8527 \\
 &                 & USFM            & 0.8156 & 0.9385 & 0.8096 & 0.7807 & 0.7815 & 0.9372 \\
 &                 & FetalCLIP       & \textbf{0.9756} & \textbf{0.9868} & \textbf{0.9770} & \textbf{0.9700} & \textbf{0.9731} & \textbf{0.9916} \\
\bottomrule
\end{tabularx}
}
\vspace{-3mm}
\end{table*}

\subsection{Evaluation Metrics}

Performance was quantified using accuracy, Area Under the receiver operating characteristic Curve (AUC), precision, recall, F1-score, and specificity. For each model, metrics were computed per class and then averaged to report macro performance, ensuring balanced evaluation regardless of dataset imbalance. 

\section{RESULTS}

\noindent\textbf{Full Fine-tuning.} Table~\ref{tab:benchmark_results} summarizes the performance under full fine-tuning. On the Spanish test set, USFM achieved the highest overall performance across all metrics (ACC = 0.9521, AUC = 0.9962, F1 = 0.9476), followed closely by natural image pretrained CNN models. ViT-based DINOv3 performed notably worse. A similar trend was observed in the out-of-domain evaluation. When tested on the African dataset, USFM maintained strong performance (ACC = 0.9556, AUC = 0.9938, F1=0.9515), demonstrating superior generalization across populations. MOFO and CNN-based architectures showed moderate performance drops, while DINOv3 and UltraSAM exhibited the most severe degradation under distribution shift.

\noindent\textbf{Linear Probing.} Table~\ref{tab:benchmark_results_frozen} reports the linear probing results for all models. On the Spanish in-domain test set, FetalCLIP was the top-performing model, achieving the highest scores across all evaluation metrics (ACC = 0.9321, AUC = 0.9938, F1 = 0.9261). It was followed by USFM, with natural image pretrained models ranking next. MOFO performed notably lower, and UltraSAM produced the worst results.
On the African out-of-domain dataset, FetalCLIP again ranked first, showing strong generalization with the best overall performance (ACC = 0.9756, AUC = 0.9868, F1 = 0.9731). USFM was again the next best-performing model, followed by natural image pretrained models, with EfficientNetV2 performing the best among the group. As in the in-domain evaluation, UltraSAM recorded the lowest performance.

\section{Discussion}

FetalCLIP was the best performing model under the linear probing setting. Its pretrained representations are already highly aligned with fetal ultrasound anatomy and texture patterns, learned from a vast multimodal ultrasound dataset. Freezing the encoder preserves these strong domain-specific features, allowing the classifier to map them directly to the target labels. In contrast, full fine-tuning updates the entire encoder, which can overwrite these high-quality features and lead to overfitting when trained on a smaller labeled dataset, as seen in our results. 

USFM was the best performing model in the full fine-tuning setting and the second best under linear probing. Its large-scale MIM pretraining on over 2.1 million ultrasound images produces broad, flexible representations that are not tied to any specific organ or task. These low- and mid-level ultrasound features remain highly adaptable, so allowing the encoder to update during fine-tuning helps the model reshape these general priors into more discriminative features for fetal plane recognition. This adaptability makes USFM particularly responsive to task-specific optimization, leading to stronger results when the encoder is fully trainable. 

CNN-based models performed better in the full fine-tuning setting because their pretrained feature spaces are not inherently aligned with ultrasound appearance. Allowing the encoder to update lets it adapt to modality-specific characteristics, which linear probing cannot when the encoder is frozen. 

%reduce DinoV3 and MOFO

DINOv3 performs better on the Spanish dataset when frozen because keeping the natural-image pretrained encoder fixed prevents overfitting. Once fully fine-tuned, these features drift and hurt in-domain performance. However, even in the frozen setting, DINOv3 still generalizes poorly to the African dataset because its natural-image representations do not capture ultrasound-specific structure.

MOFO outperforms UltraSAM, even with far fewer training images (7,039 vs.\ 280,000), because its pretraining focuses on organ-level semantic features through task prompts and anatomical priors, making dataset size less determinant. UltraSAM instead learns segmentation-oriented representations that emphasize boundaries rather than the global appearance cues needed for plane classification. Both models perform poorly under linear probing because freezing the encoder traps them in this segmentation-focused feature space.

\section{Conclusions}

%FM models are the best, it's not only the size. MIM and CLIP., more inherently try to more general about  the data. Self-supervised, non-task. 
%You chose your FM ccording to the resources, (GPU availability and how large labeled dataset you have). If you dont have  a lot of resources, use Fetal CLIP and if you have more resources for full fine-tuning, then use USFM. 

Our results show that ultrasound-specific foundation models can provide meaningful advantages for fetal plane classification, though this benefit is not uniform across all architectures. Among the four ultrasound FMs we evaluated, FetalCLIP achieved the strongest performance in the linear probing setting, while USFM performed best under full fine-tuning. This pattern reflects their underlying objectives: CLIP-based and MIM-based models learn broad, transferable representations, whereas segmentation-oriented models (UltraSAM, MOFO) learn more task-specific structural cues that do not generalize as effectively to classification tasks, particularly when the encoder is frozen.

In practice, we recommend selecting the FM based on computational resources and the size of the labeled dataset. For low-resource scenarios or small labeled datasets, linear probing with FetalCLIP is the most effective option. When full fine-tuning is feasible, USFM provides the strongest performance among the fully trainable models.
 
\section{Compliance with Ethical Standards}
This research study was conducted retrospectively using human subject data made available in open access by Burgos-Artizzu et al.~\cite{burgos2020evaluation} and Sendra-Balcells et al.~\cite{sendra2023generalisability}. Ethical approval was not required as confirmed by the license attached to the open-access data. 

\section{Acknowledgments}
\label{sec:acknowledgments}
The authors thank Rui Wang, Jiarui Xing,  and Yinchi Zhou for helpful discussions. No funding was received for conducting this study. The authors have no relevant financial or non-financial interests to disclose.

\bibliographystyle{IEEEbib}
\bibliography{ISBI2026_US}

\end{document}